\documentclass[preprint,tightenlines,eqsecnum,aps,epsfig,floatfix]{revtex4}
\usepackage{epsfig}

\def\beq{\begin{equation}}   \def\eeq{
\end{equation}}

{\large
{\LARGE }}\begin{document}
\title{ The
causality
and/or energy-momentum  conservation constraints on
QCD amplitudes in small x  regime. }
\author{ B. Blok\email{E-mail: blok@physics.technion.ac.il} }
\affiliation{Department of Physics, Technion---Israel Institute of
Technology, 32000 Haifa, Israel}
\author{ L. Frankfurt\email{E-mail: frankfur@lev.tau.ac.il} }
\affiliation{School of Physics and Astronomy, Raymond and Beverly Sackler
Faculty of Exact Sciences,
Tel Aviv University, 69978 Tel Aviv,
Israel}

\thispagestyle{empty}

\begin{abstract}
The causality and/or the energy-momentum conservation constraints
on the amplitudes of high energy processes are generalized to QCD.
The constraints imply that the energetic parton may experience at
most one inelastic collision (and arbitrary number of elastic
collisions). and  that the number of the constituents in the light
cone  wave function of the projectile is increasing with the
collision energy and the atomic number.
 \end{abstract}
\maketitle

\setcounter{page}{1}
\section{Introduction}

\par
The high energy behavior of the QCD amplitudes has attracted a lot of
attention, both experimental and theoretical. This interest is  focused on
the new QCD phenomena and on a necessity to evaluate reliably the
QCD effects accompanying the new particles production. The
aim of the present paper is to study the role of the
causality and the energy-momentum conservation in the particle
production at high energies in the perturbative QCD.
\par These constraints are absent in a relativistic quantum mechanics,
and appear only in a Quantum Field Theory, in particular in the perturbative
QCD (pQCD). In this paper  we establish the constraints imposed on
the high energy scattering amplitudes in the hard QCD by  the
causality and by the energy-momentum conservation.  The closely
related constraints were studied in a detail before the advent of
QCD, in the framework of the Reggeon Calculus and $\phi^3$
theories, see refs.  \cite{Gribovbook,ChengWu1,mandelstam1,mandelstam2}
and references therein. However, these constraints were mostly
put aside afterwards, since the dominance of the leading twist (LT)
approximation in hard processes at moderately small Bjorken x had
made unnecessary
investigation of multiple scattering processes.
\par
It has been understood in literature, for the review and
appropriate references see ref. \cite{FSW}, that the rapid
increase with the energy of the leading twist perturbative QCD
(pQCD) amplitudes
\cite{Gross-Wilczek,Dokshitzer,BFKL,LF,CC,Altarelli,Ciafaloni}
leads to the problems with the probability conservation for the
leading twist approximation in the kinematics covered by LHC and,
probably, by  the leading parton production at RHIC. Thus it seems
necessary to develop an adequate theoretical treatment of the pQCD
regime of strong interactions with small running coupling constant
to  resolve the problem  with the violation of the probability
conservation in the LT approximation at high energies.
Recently there was a number of attempts to generalize the Reggeon
Field Theory to pQCD \cite{BF,Kovner,Geller,Mueller2}.
\par
In the present paper we explore the decrease of the pQCD amplitude
with the vacuum quantum numbers in the crossed channel with the
virtuality of the parton.  This observation helps to generalize to
pQCD the famous S.Mandelstam-V.Gribov
\cite{mandelstam1,mandelstam2,Gribovbook}  proof of the
cancellation of the contribution of planar diagrams  into the
total cross section. The eikonal graphs due to the s channel
iteration of the color singlet ladder exchanges form a subset of
planar graphs. Hence their contribution is cancelled out.
\par
The complimentary constraints follow from the analysis of the
multi-particle cross-sections as  determined by the s channel
cuts of the multi-ladder diagrams. We shall show that the account
of the energy-momentum conservation leads to the complimentary
explanation why the
contributions of the
planar (in particular eikonal) diagrams are zero in the perturbative
QCD at high energies.
\par
In order to visualize constraints derived in the paper it is
rather convenient to choose the reference frame where the
projectile is energetic but the target is at rest.
\par
The qualitative explanation of the cancellation of the eikonal
diagrams due to the causality is that the projectile fragments
into the number  of the particles that is increasing with the energy. These
particles have no time to form back an incident particle in the intermediate
states due to the Lorenz dilatation \cite{Gribovbook}.
\par
We have mentioned above that the reasoning used in the paper
applies only to the iteration of the ladders with the singlet
color quantum numbers. In the case of the s-channel iteration of
ladders with color octet quantum numbers, the constraints
discussed above are inapplicable (see the discussion in chapter
2). The amplitudes with the octet quantum numbers $8_F$ in the
crossed channel do not decrease with the
parton
virtuality since
they have
the strongly off shell  intermediate state in the corresponding
Feynman diagrams
relevant for  the gluon reggeization in the pQCD. So the
contribution of the two gluon exchange in the negative signature
is absent in the eikonal approximation where all particles are on
mass shell. At the same time eikonal diagrams where "potential" is
given by reggeized gluon exchange are not forbidden by constraints
discussed in the paper.
The reasoning based on the energy-momentum conservation is also
inapplicable because the intercept  of the reggeized gluon is less
than 1 and the amplitudes are predominantely real.

\par
Let us stress here once again that our results apply only to the
quantum field theory at sufficiently high energies, but not to the
quantum mechanics or the low energy field theory. The main
difference between the quantum mechanics and the quantum field
theory is in the absence of particle creation. In the quantum
mechanics one
iterates the predominantly  real amplitudes
arising due to the
one particle (one-photon, one -gluon) exchange
and obtain s
the solution of the problem : the high energy
projectile scattering off the static center.
On the contrary, in the high energy quantum field theory one has
to iterate not the one-particle exchange diagrams, but the ladders
in order to obtain the leading order (LO)  and the next to leading
order (NLO) terms that are logarithmically increasing with the
energy. These terms are predominantly imaginary. It is in this
case that the eikonal expansion fails. Formally, the reason why
the eikonal expansion fails in the field theory is that the QFT
amplitudes decrease with the invariant masses more rapidly than
$1/M^2$, contrary to the relativistic quantum mechanics where the
potential does not depend on the virtuality.
\par
It follows from the above discussion that the  sufficiently
energetic projectile parton may undergo at most one inelastic
collision, thus strongly constraining the Feynman diagrams
relevant for the high energy processes. This observation  is
particularly important for resolving the challenge with the
probability conservation in the small x processes. In particular,
we show that the wave function of the energetic parton relevant
for the n-ladder exchange must contain at least n constituents
(contrary to the eikonal approximation, where a number of
constituents is always one).
\par
The cancellation of the contribution of the eikonal diagrams to
the total cross-section has been found also in refs.
\cite{BartelsRyskin,BartelsVacca,Lipatov,DanilovLipatov}.
These papers were focused on the
generalization
of the AGK cutting rules \cite{AGK}
to
the pQCD and did not analyze the constraints due to the causality
and the energy-momentum conservation, in particular, because  the
LO BFKL approximation does not respect the energy-momentum
conservation.
\par
We find that the non-planar diagrams that take into account the
bremstrahlung in the initial state and diffraction in the intermediate
states dominate in the impact factors.
\par
The organization of the paper is the following. In the second
chapter we explain  how the account of  the causality leads to the
cancellation of the planar (eikonal) diagrams in QCD.  In the third
chapter we discuss the constraints on the Feynman diagrams due to
the energy- momentum conservation. In particular we explain that
all the s channel cuts of  the eikonal diagrams are zero.  Thus
the nonzero contributions to the multi-ladder exchange of the
amplitudes arise entirely due to the non-planar diagrams for the
impact factor ( the Mandelstam cross diagrams). In the fourth
chapter we discuss the generalizations of the Mandelstam cross
diagrams .
The conclusions are given in the section 5. In appendices we
review some known properties of the high energy amplitudes, that
are rarely discussed in the literature, in order to make the paper
self-contained. In appendix A we remind the reader  the
Mandelstam-Gribov arguments and their derivation in the field
theory . In appendix B we review briefly the definition and the
properties of the impact-factors.
\section{The planar diagrams and
the causality for the high energy processes.}

Historically, the eikonal  approximation in the theoretical description of
the photon(hadron)-nucleus collisions at large energies the eikonal
(the Gribov-Glauber ) approximation
is the one of the most successful phenomenological approaches
\cite{Glauber,Gribovshadowing,Iennie}. The eikonal approximation
gives the legitimate solution of the Schrodinger, Dirac and
Klein-Gordon equations describing the interaction of  a
sufficiently energetic particle with a
statical source.
However, in the beginning of sixties, long before the advent of QCD,
it has been understood  that the exchange of double Pomerons,
even  with the intercept $\alpha(t=0)\ge 1$,
rapidly decreases with the energy if the impact
factors are dominated by the planar (and in particular, eikonal )
diagrams, (see Fig. 1). This result follows from the analytic
properties of the amplitude (causality) and decrease of the
amplitude  with the  virtuality of a colliding particle
\cite{mandelstam1,mandelstam2,Gribovbook}. Thus the eikonal
approximation, while is useful tool in the
quantum mechanics fails in the Quantum Field Theory.
\par
Under the influence of the evident phenomenological success of the
eikonal approximation in the description of the hadron -nucleus
collisions  this approximation is often used in gauge theories as
an attempt to cure the rapid increase of the amplitudes with the
energy and to restore the probability conservation ( see e.g.
refs. \cite{Kowalski,Sjostrand}
for a review of some recent
eikonal-based models in QCD). The standard  form of the eikonal
models is \cite{LLQ,ChengWu1,ChengWu2,ChengYan,Mueller} :
\begin{eqnarray}
\sigma_T&=&2\int d^2b (1-\exp(-a(s,b)))\nonumber\\[10pt]
\sigma_E&=&\int d^2b (1-\exp(-a(s,b)))^2\nonumber\\[10pt]
\sigma_{inel}&=&\int d^2b(1-\exp(-2a(s,b))),\nonumber\\[10pt]
\label{add}
\end{eqnarray}
where $\sigma_T,\sigma_E,\sigma_{\rm inel}$ are the the total, elastic
and inelastic cross-sections respectively, and $a(s,b)$ is an
eikonal phase  for a given impact parameter $\vec b$, calculated
perturbatively.

\par
The aim of this section is to show that the eikonal approximation
breaks down in QCD for sufficiently high energies . We show that
the eikonal iterations of the amplitude with the vacuum quantum
numbers in the crossed channel (e.g. the  BFKL,DGLAP ladders)
rapidly decrease with the  energy. We discuss two complimentary
reasons for the cancellation of the planar diagrams: one is the
generalization to QCD of the Mandelstam
\cite{mandelstam1,mandelstam2,Gribovbook} cancellation as the
consequence of the causality, another is the impossibility to
satisfy the energy-momentum conservation constraints for the
planar (eikonal) graphs for the particle production. The dominant
contribution is given by the non-planar diagrams.
\par
The cancellation of the eikonal diagrams was first proved in the
$\phi^3$ theory and generalized to the reggeon calculus, (see
appendix A for a short review of the proof). The origin of this
cancellation is that in a field theory, contrary to the quantum
mechanics, there exists the particle creation. As a result, for
sufficiently high energies the amplitude is dominated by the
exchange of the  ladders.
The eikonal representation breaks down because a
the parton can not have more than one inelastic scattering (i.e.
one attachment to the ladder exchange). More intuitively, the
ladder creation means the creation of a large number of particles,
while the two ladder iteration in the QFT means these particles
after being created, then come together into the same configuration
after finite time. This clearly looks
implausible.
\par
In this section we somewhat generalize the causality
(Mandelstam-Gribov) reasoning explaining such a cancellation.
\par
Let us  extend the Mandelstam-Gribov argument to QCD. Our starting
point is a two body collision at high energies  as given by the
single-ladder exchange .  Such construction arises in the LO and
NLO logarithmic approximations in pQCD. In this case it is
legitimate to neglect the longitudinal momentum transfer and the
denominators in the propagators in the ladder are $\sim r^2_t$
where r is the momentum transverse to momenta of colliding
particles in the line of the ladder (see Fig. 2). Then the
calculation leads to the simple form for the collision amplitude
A(s,t) \cite{ChengWu1,GLF}:
 \beq
 A(s,t)=\int d^2k d^2k' \Phi_1(p_A,k,q-k)f(s,k,q-k,k',k'-q) \Phi_2(p_B,k',q-k'),
 \label{63}
 \eeq
where f depends in pQCD on the s-channel energy squared s, and on
the transverse momenta $k,k-q,k',k'-q$, $t=-q^2_t$. The
$\Phi_{1,2}$ are the impact factors describing   the upper and
lower blobs in the Feynman diagrams of fig. 2. These impact
factors (see Appendix B for a short review of the impact factor
formalism)
as the consequence of the dominance of the
single
gluon polarization in the propagator of the exchanged gluon
\cite{Gribovbook} have the form:
\beq
\Phi_1= \int d M^2 (1/s^2) p_B^\mu p_B^\nu
f_{\mu\nu}=\int d M^2 k_t^\mu (q-k)_t^\nu f_{\mu\nu}/(M^2)^2
\label{64}
\eeq
where $M^2$ is the square of the mass of the diffractively produced
state. In the last equation one uses the Ward identities and the two
body kinematics. We use the Sudakov parametrization for the
momentum of the exchanged gluon: $k=\alpha p_A+\beta p_B+k_t$.
If the function $f$ has the form of the Regge pole exchange:
\beq
f\sim s^{\alpha(-q_t^2)}F(k_t,k' _t,q_t),
\label{65}
\eeq
the amplitude acquires the form:
\beq
A(s,t)\sim s^\alpha (t)G(t).
\label{66}
\eeq
In the pQCD the leading singularity is the Regge cut, not a pole as in  the
$\phi^3$ or the Reggeon field theory. Nevertheless, the one "ladder"
exchange amplitude can be written as
\beq
A(s,t)=R(t) \kappa (s,t)
\label{67}
\eeq
where
\beq
R(t)=\int ds_{12}\int ds_{34}d^2k_t d^2k'_t
\Phi_1(s_{12},k_t,q_t-k_t) F(k_t,k'_t,q_t)
\Phi_2(s_{34},k'_t,k'_t-q_t).
\label{68}
\eeq
Here $\kappa(s,t)$ is the function calculable in pQCD. The dependence on
the momenta $k_t$ running in the ladder is factorized in the
Particle-Particle-Reggeon ( PPR) vertex \cite{lipatov,Ciafaloni}.
\par
Let us now proceed to the evaluation of two ladder exchange (Fig.3).  The
general expression for the 2-"Reggeon" exchange amplitude can be
obtained similarly:
\begin{eqnarray}
A(s,t)&=&\int d^2r_t d^2k_t d^2r'  d^2u d^2u' d\alpha s d\beta s
\Phi_1(p_A,r,k,q,u) \Phi_2(p_B,r',k',q,u')\nonumber\\[10pt]
&\times&f_1(r,r',k,s_1)f_2(q-k,u,u',s_2).\nonumber\\[10pt]
\label{69}
\end{eqnarray}
Here $f_1$ and $f_2$ are the functions corresponding to the two
exchanged ladders, $s_1$ and $s_2$ are their invariant energy
squared in the s-channel. For the eikonal diagrams $s_1=s_2=s$
while for the Mandelstam cross $s_1, s_2\le s$, and
$\sqrt{s_1}+\sqrt{s_2}=\sqrt{s}$ (see below). The vectors $r,u,r',u'$ are
the momenta that
propagate
through these 2 ladders. The impact factors $\Phi_1,\Phi_2$ correspond
now  to the 6-point blobs. We made as above a Sudakov expansion:
$u=\alpha_u p_A+\beta_up_B+u_t,r=\alpha_r p_A+\beta_r p_B+r_t,
 k=\alpha_k p_A+\beta_k p_B+k_t$.

The invariant masses are $s_{\beta=1,2}=(p_A+r)^2=\alpha_r
s_1,s_{34}=(p_A+u)^2=\alpha_u s_2, (p_A+k)^2=\alpha s$. In the two
body kinematics the integration $d^4k$  factorizes between the
upper and the lower blobs, like in $\phi^3$ theory. Using
$d\alpha= ds_a/s, d\beta = ds_b/s$ we obtain that the amplitude of
the two-reggeon exchange is given by
\begin{eqnarray}
A(s,t)&=&\int d^2r_t  d^2u_t d^2k_t\int ds_a \Phi_1(p_A,u,r,k,q,s_a)\nonumber\\[10pt]
&\times& \int d^2r' d^2 u' d^2 k'
\int ds_b\Phi_2(p_B,u',r',k',q,s_b)f_1(r,r',k,s_1)f_2(q-k,u,u',s_2)\nonumber\\[10pt]
\label{71a}
\end{eqnarray}

For high energies the functions $f_i$ have the form of the product
of $s_{1,2}$ in some power and the function depending only on the
transverse components of the vectors. In particular, the
dependence on the invariant masses $s_{12},s_{34},s_a$ and
$s'_{12},s'_{34},s_b$ is factorized between the blobs like in the
case of the $\phi^3$ theory. Then we can use the Mandelstam
reasoning (see appendix A) . For example, the integral over $s_a$
still has the same analytical properties as for the 4-point blob.
The corresponding integration contour is depicted in Fig. 4 and is
the same for the QCD and for $\phi^3$ theory. Then one can deform
the contour of the integration into the complex plane due to the
absence of the left cut for for arbitrary planar diagrams for whom the
spectral density $\rho_{s,u}$ in the Mandelstam representation of the
amplitude $\rho_{s,u}$ is zero (in particular for eikonal diagrams). The
only remaining point is that in the eikonal diagram we have $1/s_a$
dependence due to a single particle exchange. The additional
dependence on $1/s_a\equiv 1/M^2$  of the blob follows from the
dependence of the ladder on the invariant mass.
\par
In the case of the Reggeon Field Theory this dependence is
$\sim 1/(M^2)^{\alpha(t}$ with $\alpha(0) \ge 1$ \cite{FS89},
meaning that the
impact factor
decreases faster then $1/s_a$ as a function of the invariant mass .
The similar dependence on the inavariant mass is valid  in the
perturbative QCD for the exchange of the color singlet ladder.
\par
Indeed, let us consider the color singlet pQCD ladder amplitude
for the case  of two different invariant masses $Q^2$ and $M^2$.
This amplitude is the weak function of $M^2$ in the area $M^2\sim
Q^2$. However , if $M^2>>Q^2$ this amplitude decreases with $M^2$.
In particular, amplitude of DIS evaluated at small x within the
DGLAP approximation is $$\sim 1/(Q^2+M^2)^n$$ where $n\sim 3/4$
\cite{AFS}. Similar behavior is expected for color singlet
Generalized Parton Distributions (GPD)
within the DGLAP approximation.
 Remember that at the
achievable small $x$ and $Q^2$ pQCD amplitudes evaluated within
the NLO DGLAP and NLO BFKL approximations are rather
close\cite{Altarelli,Ciafaloni}.
\par
In contrast to the pre QCD approaches which conveniently assumed
the fast decrease of any  amplitudes with $M^2$, the pQCD gives
two different patterns. The amplitude with the color octet quantum
numbers in the crossed channel does not depend on $M^2$ for
$s\ge M^2$ at least within the leading logarithmic  approximation.
This is because the virtuality of the interacting parton in the ladder
is $\le s$.
Therefore such an amplitude does not decrease with the virtuality
of a parton. Besides the color octet amplitude is predominantly
real and decreases with the energy, so the analysis of the
multiparticle states through s- channel  cuts is unreliable
in this case.
\par
The above proof also makes clear why the eikonal expansion is
valid in quantum mechanics, as it was mentioned in the introduction.
Indeed, for the one particle exchange, contrary to the ladder exchange,
the impact factor decreases like $1/M^2$ at most, and the contour of fig. 4
can not be deformed, even if the left cut is absent.
\par
In this paper we, however, are interested in the
amplitude with the vacuum quantum numbers, where the virtuality of the
interacting parton is $\approx  \sqrt Q_1^2$. Here $Q_1^2$ is the
order of the  maximum between $Q^2$ and $M^2$ . In this kinematics,
$Q^2\ll M^2$, $Q_1^2\sim M^2$. Then the amplitude with the vacuum
quantum numbers in the crossed channel for the scattering of a
parton is  approximately proportional to $(1/M^2) S(M^2/Q_1^2)$,
where S is the Sudakov form factor. Similar dependence is valid
for the amplitude for the scattering of the dipole. Thus we have
an additional $M^2$ dependence for the eikonal diagrams. Such a
decrease is sufficient to justify the deformation of the contour of
integration in Fig. 4 in the case of the planar diagrams.
We have proved that in the multiRegge kinematics the 2-reggeon eikonal
exchange amplitude is zero. The same arguments can be used for the
multi- eikonal exchanges in the multiregge kinematics.
\par
Note that in the proof it was essential to use Ward identities
$k^\mu f_\mu=0$ for the impact factor. It is known \cite{t'Hooft} that
such form of the Ward identities is
valid
for the amplitudes where only
one gluon is off mass shell, i.e. for a sum of all permutations of
the gluon lines in the dipole. We also expect Ward identities to hold
when only color singlet exchanges are considered, as in eikonal
diagrams (see i.e. Fig 1).
\par We can  estimate the range of energies where the reasoning
discussed in this paper applies. In fact, there are several
relevant QCD regimes , depending on the problem under
consideration.
\par
In the deep inelastic scattering the one ladder contribution is
dominant in the whole kinematical region of x. The one-ladder
(leading twist) contribution breaks down at $x\sim 10^{-5}$
\cite{FS2000}, and the multi-ladder contributions become
dominant. The area where the eikonal models fall under scrutiny
is somewhere in the middle of this interval, i.e. $x\sim 10^{-3}$,
where the multi-ladder contributions first appear. Our
results show that the eikonal contribution should be zero.

\section{Constraints due to the energy-momentum
conservation.}
\par
In the previous chapter we  concluded that the eikonal contribution is zero if
all the ladders are color singlets. In this chapter we shall argue, that these
results can be derived also from the requirement of  the energy-momentum
conservation.  We shall first review the constraints due to energy-momentum
conservation in $\phi^3$ theory and then extend this reasoning to QCD.
We shall see that the constraints due to energy momentum conservation
are sometimes even stronger than the Mandelstam one leading to the
cancellation not only of the eikonal diagrams but also of all
cut eikonal diagrams.
\par
Let us start first from the $\phi^3$ theory ,from the eikonal
graph of fig. 5.
This graph corresponds to the amplitude of the multi-particle
creation
resulting from the s-channel cut of the
two ladder diagram. Then a total square of the energy of the
created particles is $2s$, while the initial energy is $s$.
Thus initial parton releases in the two consequent scatterings the
double of initial energy.
Evidently as the consequence of the energy-momentum  conservation
law the contribution of this graph should be zero.
\par
Note that the
non-conservation of the energy by the cut eikonal diagrams is well
known for the hadron-nucleus collision for some time
\cite{ChengYan,Shabelsky,CapellaKaidalov}. However, while a
prescription has been suggested  how to include by hand the energy
conservation law into the eikonal diagrams \cite{CapellaKaidalov},
this suggestion has no justification in the perturbation theory
cf. discussion in ref. \cite{Iengo}.
\par
Let us recall that the s channel cut diagrams carry additional
information as compared  to the imaginary part of the total
amplitude of the two body scattering . The s channel cut
amplitudes are relevant for the multi-particle cross-sections
\cite{AGK}. Indeed, if the average number of particles created in
one reggeon exchange is $\sim \bar n$, then the contribution of
the multi-ladder exchanges leads  to the processes in which number
of particles produced is a multiple of $\bar n$. The cross-section
of the creation of $n\bar n$ particles is dominated by the
diagrams with n cut ladder exchanges.  In other words the energy
conservation law must be fulfilled separately for the diagrams
with n cut ladders for each n. However for n cut ladders initiated
by planar diagrams the square of invariant energy of the particles
created is $n s$, while the original energy is $s$, the energy
conservation law is violated and all the cut diagrams are zero.
The related reasoning is to check that the momentum sum rule is
violated \cite{AGK}.
\par
The above reasoning can be directly translated to QCD without any
changes since the imaginary parts of the ladders are significantly
larger  than the real parts in QCD, within both in the BFKL and
DGLAP approximations. This means, that the contribution of eikonal
diagrams with the  NLO  BFKL and  DGLAP ladders is zero.
\par
Let us note here that the energy-momentum conservation prohibits
radiation of a more than one inelastic ladder by a single parton.
On the other hand the parton can have an arbitrary number of
elastic rescattering on the target, since such rescatterings does
not change the energy of the energetic parton.

\section{The Mandelstam cross.}
\par
We have explained that the eikonal iteration of the ladders gives
zero in QCD. The obvious question is what is the dominant
contribution. The simplest diagram that gives  nonzero 2-ladder
contribution is Mandelstam cross diagram of Fig. 6
\cite{mandelstam2}. The contribution of this cut can be easily
derived in $\phi^3$ theory , one just takes into account that the
energy is split between the two ladders. One has \cite
{Gribovbook}
\beq
A(s,t)=(i/4)\int d^2k_tN^2_{\gamma\gamma_1}
(k_t,q_t)\xi_\gamma\xi_\gamma s^{\gamma+\gamma_1-1}
\label{71b}
\eeq
Here the squared energy parameters of the ladders are
different for 2 ladders:
\beq
s_1=\alpha s, s_2=-s\beta
\label{72b}
\eeq
and $\xi_\gamma=-(\exp(-i\pi\gamma +1)/\sin (\pi\gamma)$,
\beq
N_{\gamma\gamma_1}=\int ds_1 d^2k_t g_1g_1^1\beta_1^\gamma
(1-\beta_1)^{\gamma_1}/(.....)
\label{73}
\eeq
In the same way one can easily derive the contribution of the
Mandelstam cross in the case of QCD. The answer is
\begin{eqnarray}
A(s,t)&=&(i/4)\int \frac{d^2k_t}{(2\pi)^2}\int\frac{d^2u_t}{(2\pi)^2}\Phi_1(s,\alpha,\beta,k_t,u_t)\nonumber\\[10pt]
&\times&\Phi_2(s,\alpha,\beta,k_t,u_t)f_1(\alpha s,k_t)f_2(\beta s,u_t)\delta (\alpha +\beta -1)\nonumber\\[10pt]
\label{crocodile}
\end{eqnarray}
Here $\alpha +\beta =1$, and the energy conservation law is
fulfilled autumatically. Evidently the contribution of the
Mandelstam cross is nonzero.
Moreover,
one can not add additional ladders to
Mandelstam cross without increasing the number of constituents in
the s channel. Indeed, consider the diagrams of fig. 7. It is easy
to show using the arguments of the previous chapters that all
these diagrams are equal to zero. The simplest nonzero diagrams
with the three ladder exchange are the so called nested diagrams
\cite{Hasslacher}. These diagrams have 3 constituents in the
intermediate state.
\par
We conclude, that at high energies there is no universal impact
factor, that can be iterated. For n-ladder diagram to be nonzero,
one needs to have at least n constituents in the wave function of
the energetic projectile in the s-channel, i.e. not more than one
ladder can be attached to a given line.
\par

\section{Conclusion}

\par
We have shown that  the cancellation of the planar (in particular
eikonal) diagrams found within the Reggeon Calculus  by S.
Mandelstam is valid in QCD also  for the s channel iterations of
the color singlet ladder exchange.  As a consequence
the restriction by eikonal diagrams  leads to the decrease with
the energy of the shadowing effects
which is the artifact of the eikonal approximation.
The account of the energy momentum constraints leads to the same
conclusion. Moreover, the application of these constraints to the
analysis of the iteration in the s channel of the amplitudes
evaluated within DGLAP approximation shows that such iterations
are also decreasing as powers of energy. To obtain nonzero
exchange by n ladders incident parton should develop configuration
of n constituents long before the collision. In other words,  any
given constituent can participate in the ladder exchange at most
once.

\par
The challenging question is how to take  into account the non-planar
graphs ,in particular, that generated by the Mandelstam
crosses. At sufficiently large energies we showed that the number
of exchanged ladders and therefore the number of constituents in
the wave function of the incident particles is increasing with the
energy,  cf. ref. \cite{FSW}. The  generalization of Gribov
Reggeon Calculus\cite{Gribovbook} to the pQCD regime of strong
interaction with the small coupling constant may lead to the solution
of this problem.
\par
Our considerations show that the application of the eikonal
(Gribov-Glauber ) approximation for the dipole interactions with a
target have problems with causality and energy-momentum
conservation. As a result one needs a different approximation in
the reggeon-like approaches to the behavior of the QCD, when it
approaches the black disk limit and the LT approximation breaks
down. \acknowledgements{ We thank  L.Lipatov for the discussion of
properties of amplitudes with negative signature (color octets)}.

\appendix
 \section{ The eikonal diagrams cancellation in the $\phi^3$ theory.}
\par
In this appendix we briefly review the Mandelstam -Gribov
explanation of  the cancellation of the  planar (eikonal) graphs
in the $\phi^3$ theory. We shall follow the very transparent
derivation of this cancellation given in the Gribov's lectures
\cite{Gribovbook}, since such derivation can be directly
generalized to QCD.
\par
Let us write the expression for the
diagram 8, that describes the scattering in the $\phi^3$ theory
due to the exchange of two particles/reggeons in the t-channel.
The blob A may be, for example, the diagram that corresponds to
the eikonal interaction, a planar box diagram , or a Mandelstam
cross. Suppose that as a function of the invariant mass
$s_1=(p_1+k)^2$ the blob A decreases faster than $s_1$.  Let us
use the Sudakov variables:
\beq
k=\alpha_qp_2'+\beta_{p_1'}+k_t,
\label{31}
\eeq
\beq t=q^2=q_t^2+s\alpha_q\beta_q\sim q^2_t,
\label{32}
\eeq
\beq s_1=(p_1+k)^2\sim \alpha s, s_2=(p_2-k)^2=-\beta s.
\label{33}
\eeq
Since the amplitudes
$A(s_i)$ fall rapidly with increasing $s_i$, the essential values
of $s_{1,2}$ are of the order of $\mu^2$ where $\mu$ is the mass
of the particles. Then $\alpha,\beta\sim s$, and for the product
of particle propagators in diagram 8 we have
\begin{eqnarray}
\frac{1}{k^2-m^2}\frac{1}{(q-k)^2-m^2} &=&1/(\alpha\beta
s+k_t^2-m^2)1/((\alpha-\alpha_q)(\beta-\beta_q)s+(q-k)^2_t-m^2)\nonumber\\[10pt]
&=&1/(m^2-k^2_t)1/(m^2-(q-k)^2_t).\nonumber\\[10pt]
\label{34}
\end{eqnarray}
The full amplitude B corresponding to the diagram 8 can be
rewritten as \beq B=(i/4s) \int
(d^2k_t)/(2\pi)^21/((m^2-k^2_t)(m^2-(k-q)^2_t) \int_\Gamma
ds_1/(2\pi i)A(s_1)\int_\Gamma ds_2/(2\pi i)A(s_1) \label{36} \eeq
The integration contour is given in fig. 4. It has evidently a cut
in the s channel starting from the mass of the first intermediary
2-particle state $s=(2m)^2=4m^2$. In the negative axis the cut in
the $s_1$ plane starts from $t$. Since the function A as a
function of $s_1$ falls rapidly, we can deform the contour. The
integral will actually be zero if the left cut is absent . Indeed,
in this case we can close the integral to the left cut, and it
will be zero. As it is well known, the left cut corresponds to the
nonzero Mandelstam double spectral density $\rho_{su}(s,t)$,i.e.
\beq {\rm Im A}=(1/\pi)\int \rho_{su}(s',t)ds'/(s'-s) \label{37}
\eeq It is clear that the diagram  that corresponds to the eikonal
has no double spectral density-it is a tree diagram. In the same
way the planar diagrams have no double spectral density
$\rho_{su}$. The simplest diagram with the nonzero spectral
density that contributes to the 2-particle exchange is a
Mandelstam cross \cite{mandelstam2} of fig. 6.
\par
In quantum mechanics, or in the quantum field theory for not so
high energies, when the dominant contribution to scattering comes
from single particle exchange the blobs do not decrease with
$s_i$. So it is impossible to deform contour of integration to
$\infty$. This is why eikonal approximation is applicable in the
framework of the quantum mechanics.
\par
Let us now consider what happens when the dominant contribution to
the scattering amplitude comes from ladder exchange. For the case
of the $\phi^3$ theory S.Mandelstam and V.Gribov substituted the
one particle exchange by the reggeon . The key is that the reggeon
form factor has additional dependence on $s_1$, as it was proved
for this theory by Mandelstam \cite{mandelstam1}, and this
dependence is $1/s_1$. After  the substitution of the particles by
the ladders, due to the invariant mass decrease of the reggeon
form factors, the blob amplitude decrease now faster than $1/s_1$.
and the eikonal graph (as well as all planar diagrams for which
$\rho_{su}=0$) is zero. The two ladder contribution decreases with
the energy, contrary to the naive expectation that it rises as
$s^{2\alpha(t)}$, where $s^\alpha (t)$ corresponds to the single
ladder exchange. We conclude that the the planar (eikonal)
contribution is decreasing as a function of energy for $\phi^3$
theory.

\section{ The impact factors.}
In this appendix we shall briefly remind the reader the definition
and the properties of impact-factors. The relevant formalism was
developed by Cheng and Wu (see ref. \cite{ChengWu1} and references
therein and by Gribov, Lipatov and Frolov, \cite{GLF}, who studied
the the asymptotic behavior of the diagram of fig. 8 in QED. The main
result of refs. \cite{ChengWu1,GLF} relevant for us is that the scattering
amplitude with the exchange of vector particles of fig. 8 also factorizes
into the product of denominators of propagators in the
intermediate states and two factors (so called impact-factors)
that dependent only on the left and right blobs in the diagram
separately. For completeness
 let us summarize here their beautiful
proof. Indeed, let us once again use the Sudakov expansion for k.
Suppose the blocks that correspond to left and right blobs are
$f_1{\mu_1\mu2}$ and $f_{2\nu_1\nu_2}$. Suppose also that these
blocks do not increase as functions of invariant masses
$s_1=(p_1-k)^2,(p_2+k)^2=s_2$. Then the relevant areas of
integration are \beq s_1\sim -\alpha s+k^2_t\sim m^2,
(p_2+k)^2\sim s\beta+k^2_t\sim m^2\label{51} \eeq then for the
photon propagator denominators we see that they are equal to
$k^2_t, (q-k)^2_t$. Since the calculation is gauge invariant, we
can use the Feynman gauge and then the amplitude has the
factorized form \beq
B=-i(1/(2\pi)^4)(1/2)\int d^2k_t
\delta_{\mu_1\nu_1}\delta_{\mu_2\nu_2}\phi_{1\mu_1\mu_2}(k_t,p_1)
\phi_{2,\nu_1\nu_2}(k_t,p_2)
\label{52}
\eeq
 where
\beq
\phi_{\mu_1\mu_2}=\int_{-\infty}^{\infty}f^{1,2}_{\nu_1\nu_2}(s_{1,2},k_t,Q)
\label{53}
\eeq
Here $s_1=-s\alpha,s_2=s\beta$. Let us rewrite the
eq. \ref{52} in a more convenient form. In order to do it, we can
use the observation by Gribov, that for high energies the main
contribution comes from the so called nonsense asymptotic states.
The nonsense state of two virtual photons is a state where the
total spin projection in the direction of motion in the center of
mass (c.m.s.) reference frame of the t channel is equal to two. In
the physical region of the s channel it is possible to prove (see
ref. \cite{GLF} for details) that for the light cone components of the
vector polarization of a photon  are
\beq
e^-=p_1\sqrt{2/s},e^+=p_2\sqrt{2/s}
\label{53a}
\eeq
For each of the photons the propagator can be written as
\beq
\delta_{\mu\nu}-k_\mu  k_\nu /k^2)=e^+_\mu e^-_\nu+e^+_\nu
e^-_\nu+e^0_\mu e^0_\nu
\label{54}
\eeq
where $e^0$ is the longitudinal polarization
 vector of the photon in the t channel,
\beq
e^0=(1/\sqrt{k^2})(0,0,\vert \vec k\vert,k_0)
\label{55}
\eeq
 Then in the relevant integration area all of the external
 invariants of blocks 1 and 2 are of order $m^2$,
 i.e. $\beta_i\sim 1, \alpha_i\sim m^2/s,k^i_t\sim m$ for block 1
 and $\alpha_i\sim 1 , \beta_i\sim m^2/s, k^i_t\sim m$ for block 2.
 Hence all virtual momenta in block 1 have long components along
 the vector $p_1$ in
 the block 1 and those in 2 along the vector $p_2$. Therefore the
 largest contribution proportional to s in the equation for
 propagator will be given by the term $e^+_\mu\times e^-_\nu$ as compared
 to the term $e^0_\nu e^0_\mu\sim 1$, and $e^+_\nu e^-_\mu \sim 1/s$ :
 \beq
 \delta_{\mu\nu}-k_\mu k_\nu
/k^2)=e^+_\mu e^-_\nu\sim (2/s)p_{2\mu}p_{1\nu} \label{56} \eeq We
then obtain the explicit dependence of the amplitude B of fig.3 on
s in the almost factorized
form: \beq F=-i(s/4)\int d^2k_t
\frac{1}{k^2_t-\lambda^2} \frac{1}{(q-k)^2_t-\lambda^2}
\Phi_1(k_t,Q)\Phi_2(k_t,Q)\label{57} \eeq where
\begin{eqnarray}
\Phi_1(k_t,q)=(2/s^2)\int^\infty_{-\infty}(ds_1/(2\pi))f^1_{\mu_1\mu_2}
(s_1,k_t,q)p_{2\mu_1}p_{2\mu_2}\nonumber\\[10pt]
\Phi_2(k_t,q)=(2/s^2)\int^\infty_{-\infty}(ds_2/(2\pi))f^2_{\mu_1\mu_2}
(s_1,k_t,q)p_{2\mu_1}p_{2\mu_2}\nonumber\\[10pt]
\label{58} \end{eqnarray} We reproduced, in order to be
self-contained, the first step of the GFL derivation. In fact, it
is straightforward to obtain eq. \ref{58} even simpler, just from
the Gribov analysis
of the vector particle exchange for high
energies with the vertex $\Gamma_\mu(p,q)$ . Then it is shown in
ref. \cite{Gribovbook} that the dominant contribution to the
amplitude is due to nonsense state and then the diagram can be
written as a product of s-independent vertices, scalar particle
propagator and s. From this we can straightforward obtain the
result \ref{58}. Now we can use the Ward identities: \beq
k_{\mu_1}f^{\mu_1\mu_2}_{1,2}=k_{\mu_1}f^{\mu_1\mu_2}_{1,2}
\label{59} \eeq These Ward identities can be rewritten as
\begin{eqnarray}
 (\alpha
p_2+k_t)_{\mu_1}f^1_{\mu_1\mu_2}&=&(-\alpha
p_2+q-k_t)_{\mu_2}f^1_{\mu_1\mu_2}=0
\nonumber\\[10pt]
 (\beta
p_1+k_t)_{\nu_1}f^2_{\nu_1\nu_2}&=&(-\beta
p_2+q-k_t)_{\nu_2}f^2_{\nu_1\nu_2}=0
\nonumber\\[10pt]
\label{60} \end{eqnarray}
 Accordingly, we can rewrite the expression for the amplitude of
fig. 8 with the exchange of two vector particles in a fully factorized
form \ref{56}, with the impact factors being
\begin{eqnarray}
\Phi_1(k_t,Q)=(2/s^2)\int^\infty_{-\infty}(ds_1/(2\pi))f^1_{\mu_1\mu_2}
(s_1,k_t,Q)k_{1\mu_1}(Q_t-k_{2t\mu_21})\nonumber\\[10pt]
\Phi_2(k_t,Q)=(2/s^2)\int^\infty_{-\infty}(ds_2/(2\pi))f^2_{\nu_1\nu_2}
(s_1,k_t,Q)k_{2\nu_1}(Q-k)_{2t\nu_2}\nonumber\\[10pt]
\label{61}
\end{eqnarray}
We have the amplitude for the exchange
of vector particles and for external particles with arbitrary
spins, in the factorized
form, like in the above treatment of $\phi^3$ theory. One of the reasons we
reproduced here the main points of \cite{GLF}, was to show
that the proof is  extended without any changes into
pQCD . Indeed, all the points in the proof can be directly transferred
to QCD except two: first, the Ward identities that were used in QED are not
the same as in QCD.  In order for Ward identities
in QCD to become of practical use
at most one external line must be out of mass
surface-the condition fulfilled in the present case.
Really all particles  within ladder are on mass shell in the case
of  the amplitude of the positive signature. Therefore the cross
section is determined by the amplitudes where only one gluon is
off the mass shell.

\newpage
\begin{figure}[htbp]
\centerline{\epsfig{figure=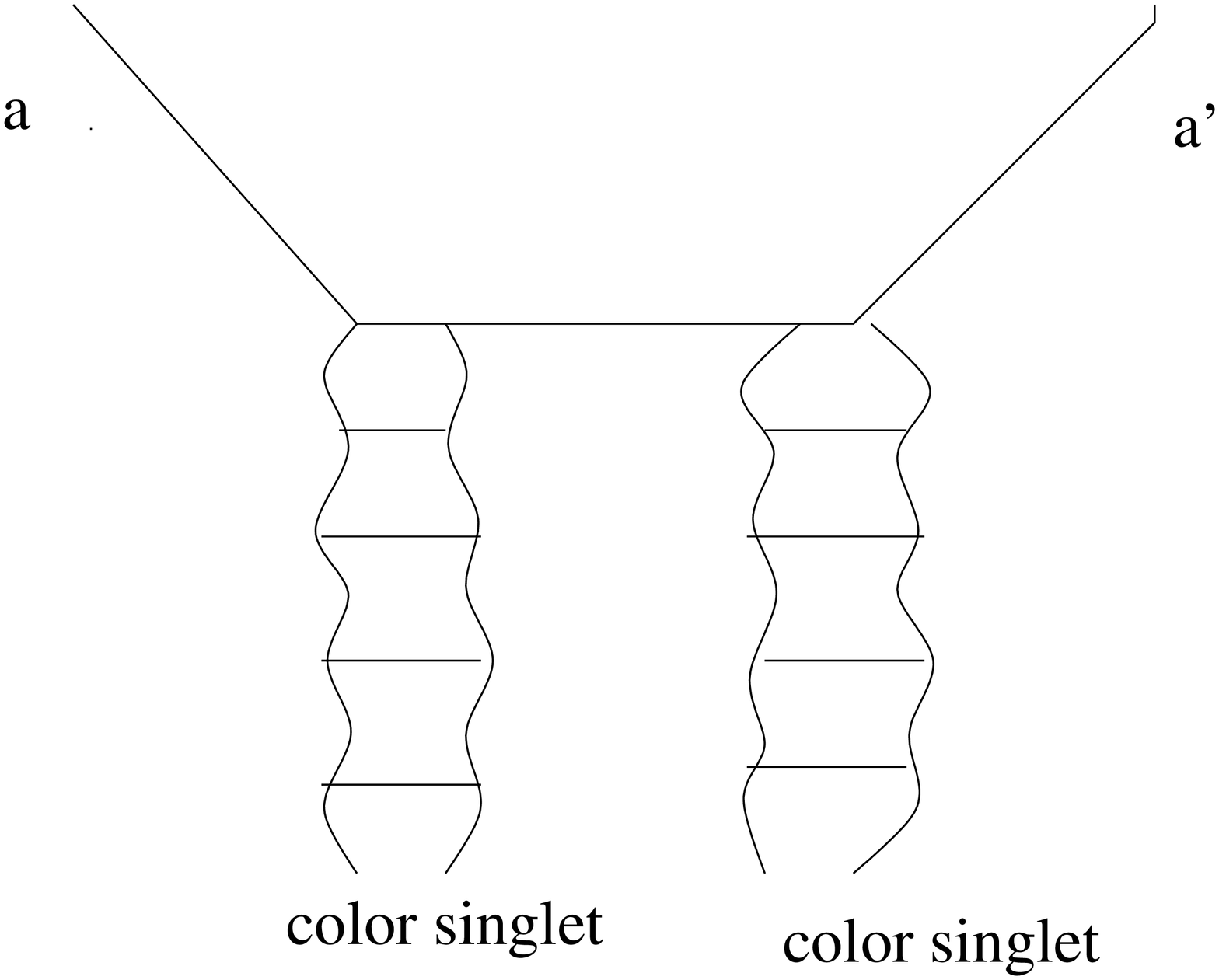,width=15cm,height=15cm,angle=-90,clip=}}
\caption{Eikonal blob for the exchange of the two color singlet
ladders in QCD} \label{fig1a.eps}
\end{figure}
\clearpage
\begin{figure}[htbp]
\centerline{\epsfig{figure=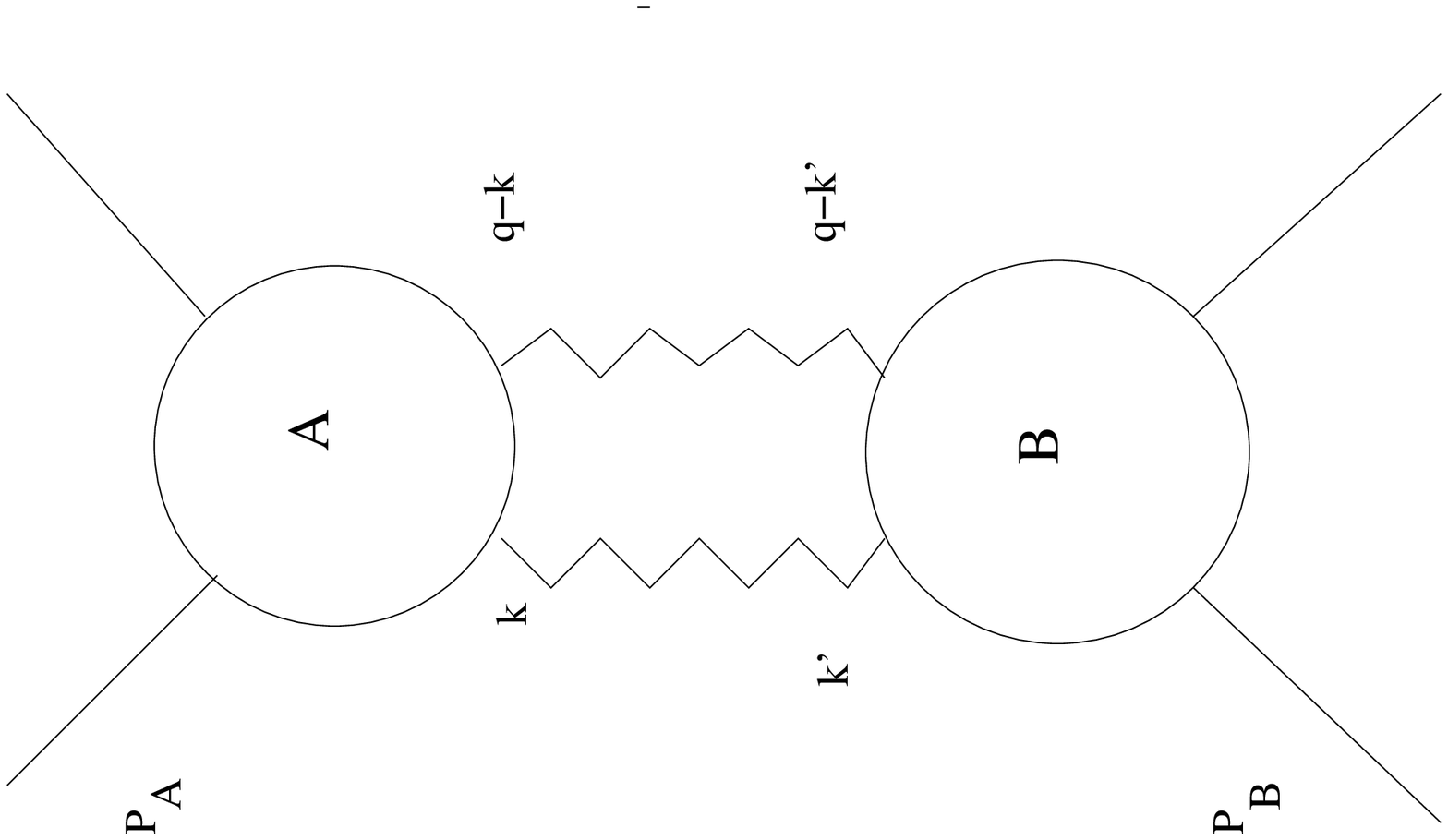,width=15cm,height=15cm,angle=-90,clip=}}
\caption{One-ladder exchange in QCD} \label{fig2a}
\end{figure}
\clearpage
\begin{figure}[htbp]
\centerline{\epsfig{figure=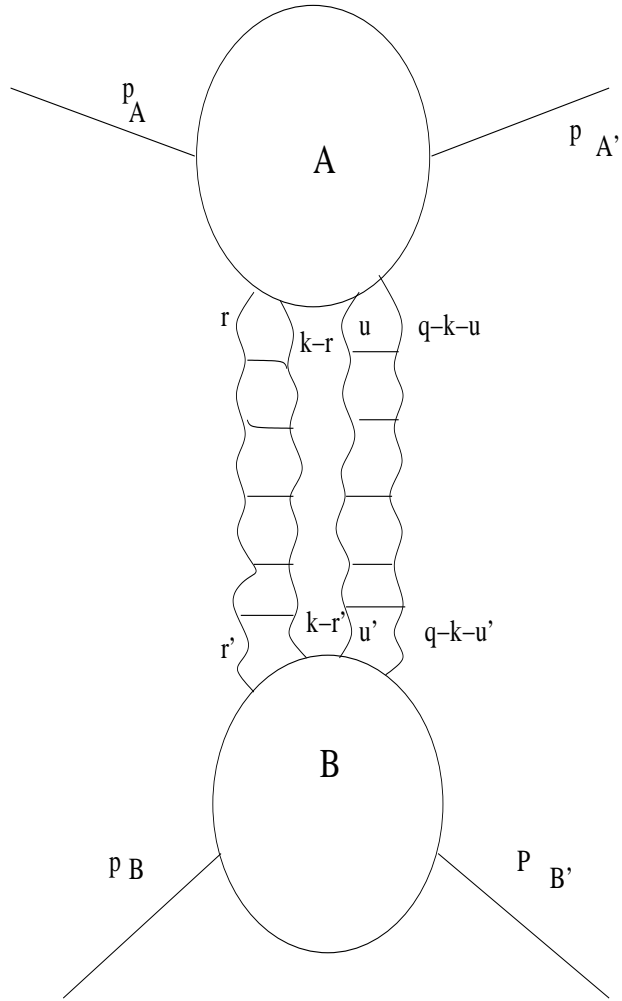,width=15cm,height=15cm,angle=-90,clip=}}
\caption{Two ladder exchange in QCD} \label{fig3a}
\end{figure}
\begin{figure}[htbp]
\centerline{\epsfig{figure=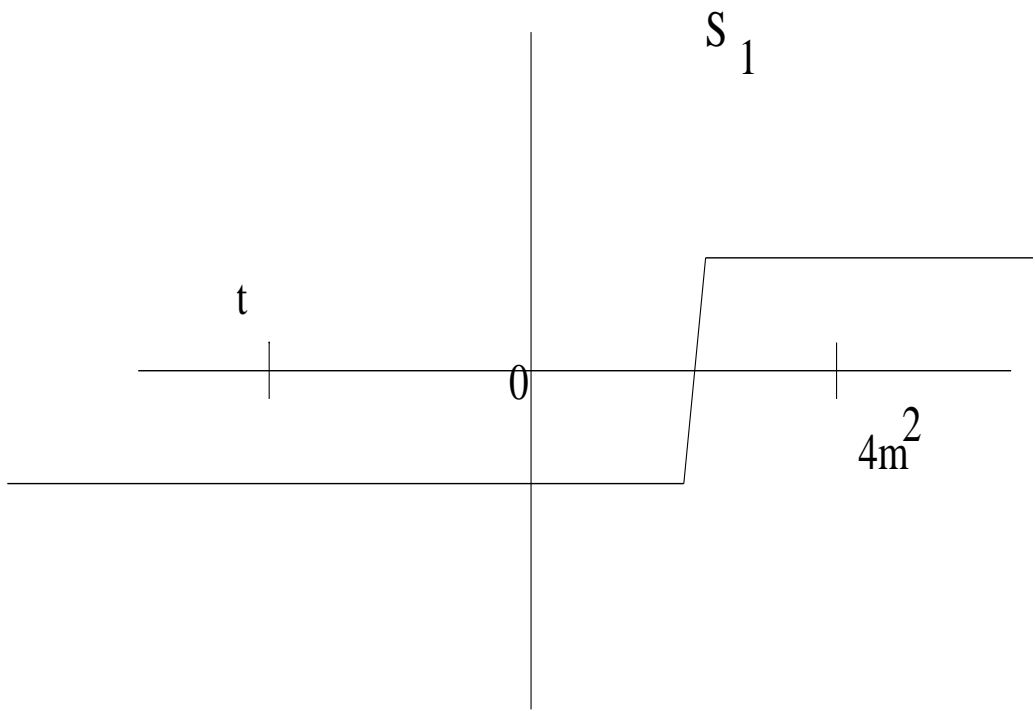,width=15cm,height=15cm,angle=-90,clip=}}
\caption{The integration over the invariant masses. For the QCD case m=0,
and the integration contour can cross the x axis also to the right from
the origin.}
\label{fig4a}
\end{figure}\clearpage
\begin{figure}[htbp]
\centerline{\epsfig{figure=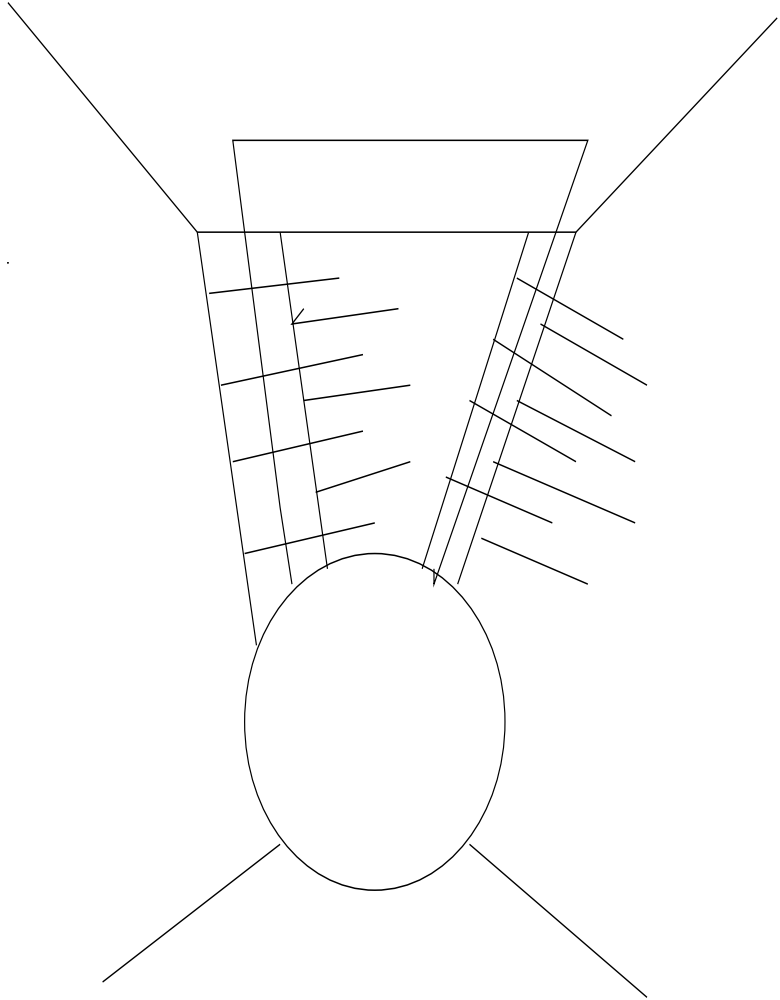,width=15cm,height=15cm,angle=-90,clip=}}
\caption{Cut eikonal graph and energy-momentum conservation.}
\label{fig5a}
\end{figure}
\clearpage
\begin{figure}[htbp]
\centerline{\epsfig{figure=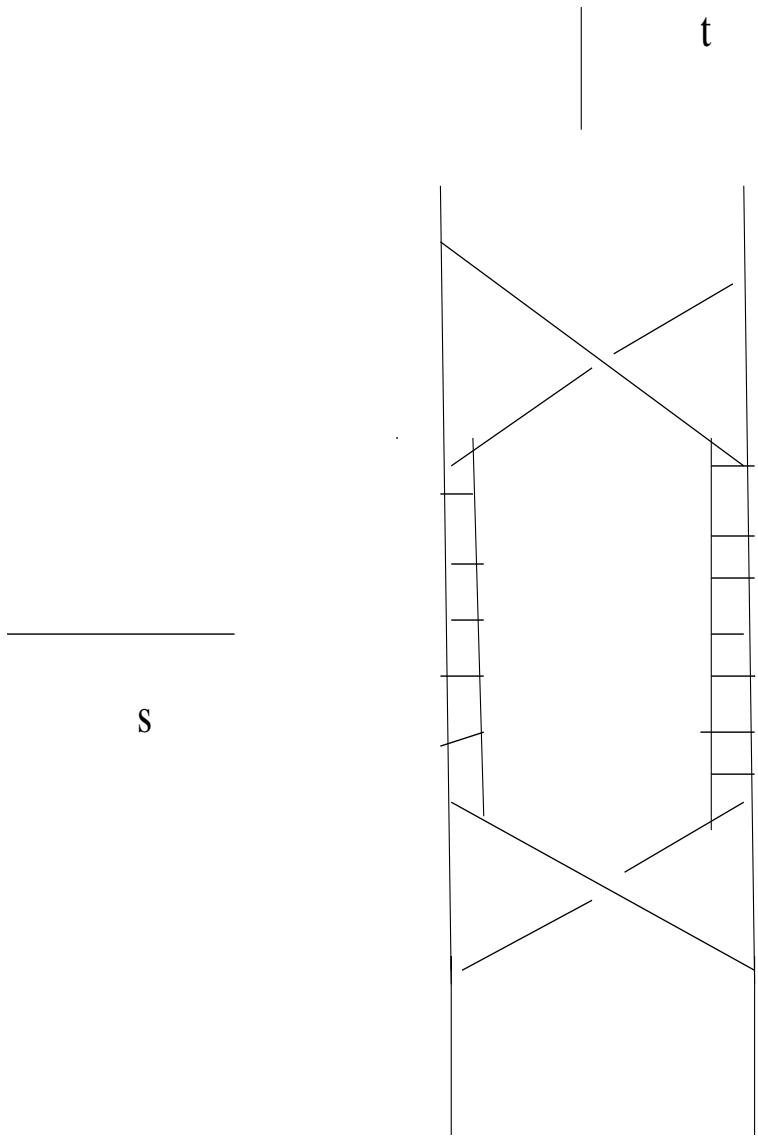,width=15cm,height=15cm,angle=-90,clip=}}
\caption{The Mandelstam cut diagram} \label{fig6a}
\end{figure}
\clearpage
\begin{figure}[htbp]
\centerline{\epsfig{figure=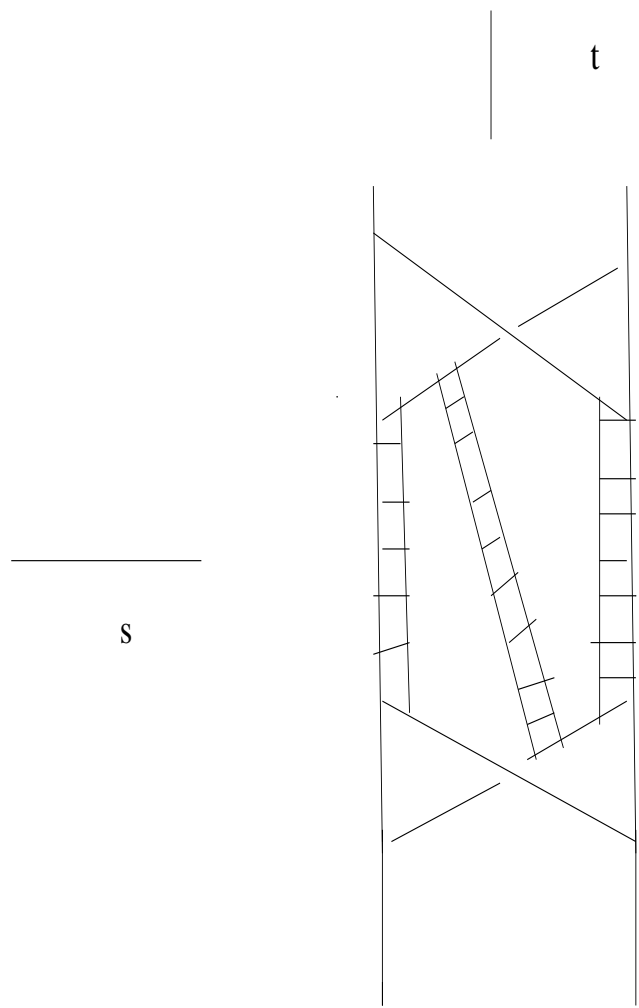,width=15cm,height=15cm,angle=-90,clip=}}
\caption{Adding ladder to  Mandelstam cut diagram}
\label{fig7a}
\end{figure}
\clearpage
\begin{figure}[htbp]
\centerline{\epsfig{figure=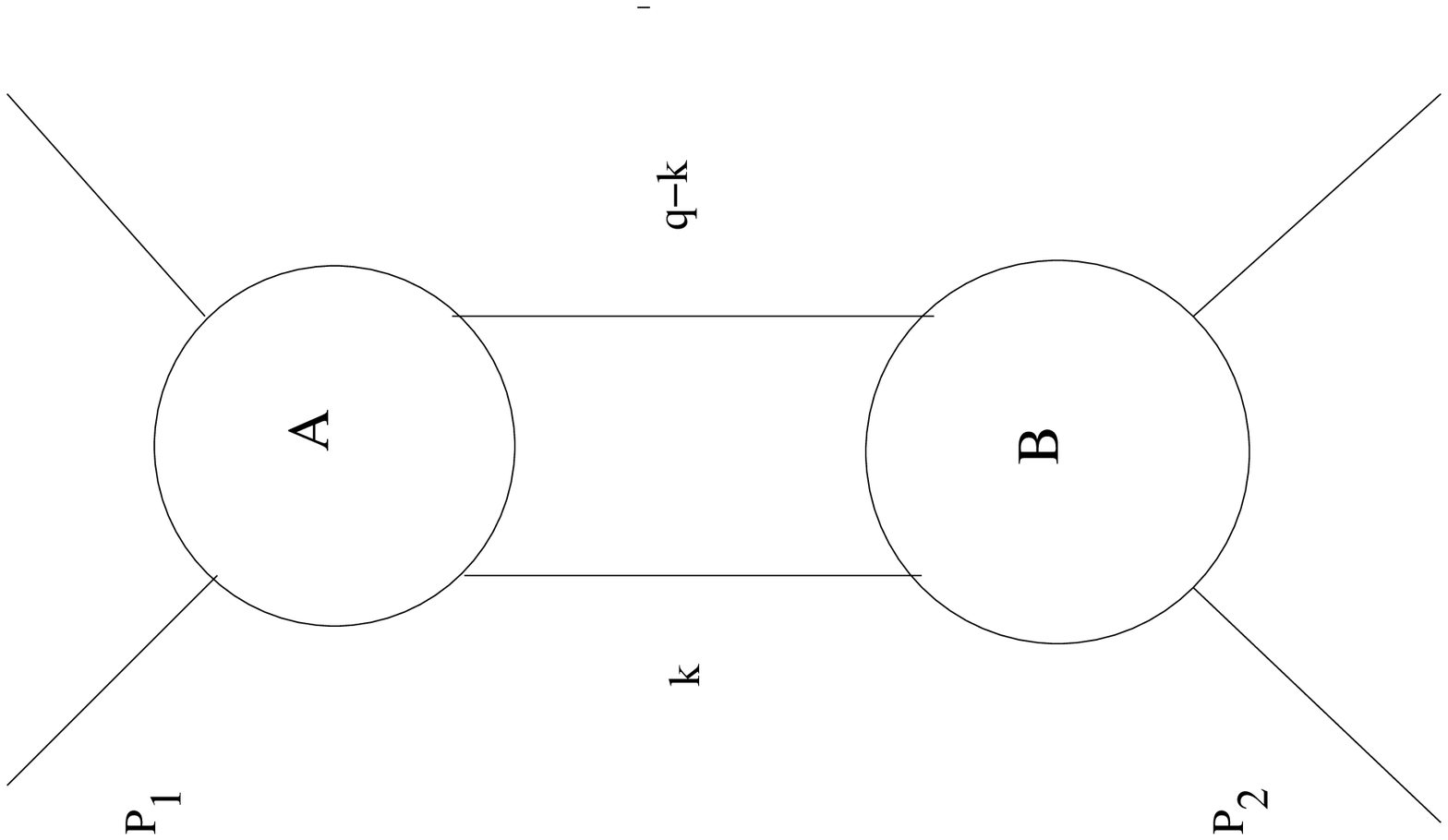,width=15cm,height=15cm,angle=-90,clip=}}
\caption{Two particle and two reggeon exchange in s-chanel in
the $\phi^3$ theory.} \label{fig8a}
\end{figure}\clearpage
\end{document}